\pdfoutput=1
\documentclass[a4paper]{PoS}
\usepackage[utf8x]{inputenc}
\usepackage{amsmath,amssymb,amsfonts}

\newcommand{\calO}{\mathcal{O}}
\newcommand{\rme}{\textrm{e}}

\renewcommand{\vec}[1]{\mathbf{#1}}

\title{Suppression of excited-state effects in lattice determination of
nucleon electromagnetic form factors}

\ShortTitle{Excited-state effects on nucleon form factors}

\author{\speaker{G.M.~von Hippel}$^a$, S.~Capitani$^a$, D.~Djukanovic$^b$,
        J.~Hua$^a$, B.~J\"ager$^c$, P.~Junnarkar$^{a,b}$, H.B.~Meyer$^{a,b}$,
        T.D.~Rae$^a$, H.~Wittig$^{a,b}$ \\
        $^a$ PRISMA Cluster of Excellence and Institut für Kernphysik,
        University of Mainz, 55099 Mainz, Germany\\
        $^b$ Helmholtz Institut Mainz,
        University of Mainz, 55099 Mainz, Germany\\
        $^c$ Department of Physics, College of Science,
        Swansea University, Swansea, SA2 8PP, UK\\
        E-mail: \email{hippel@kph.uni-mainz.de}}

\abstract{
          We study the ability of a variety of fitting techniques
          to extract the ground state matrix elements of the vector current
          from ratios of nucleon three- and two-point functions that contain
          contaminations from excited states. Extending our high-statistics
          study of nucleon form factors, we are able to demonstrate that the
          treatment of excited-state contributions in conjunction with
          approaching the physical pion mass has a significant impact on the
          $Q^2$-dependence of the form factors.\\
\vskip3ex
${}$\hfill\includegraphics[width=0.25\textwidth,keepaspectratio=]{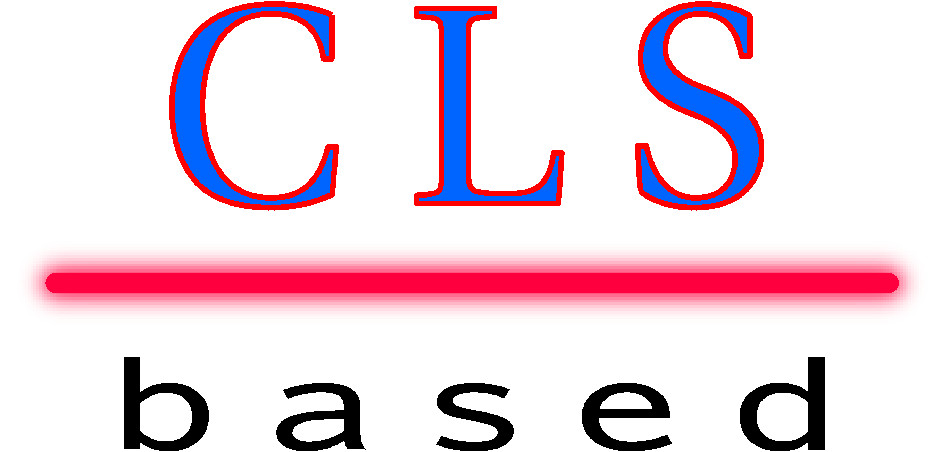}
          }

\FullConference{The 32nd International Symposium on Lattice Field Theory\\
                23-28 June, 2014\\
                Columbia University New York, NY}

\begin{document}

\section{Introduction}

\noindent The persistent failure of the vast majority of lattice measurements of nucleon form factors to replicate the experimental value for the proton charge radius is one of the biggest current puzzles in lattice QCD. Possible explanations include the use of unphysically heavy pions, finite-volume effects and excited-state contributions (or indeed a combination of some or all of these). Our purpose here is to study how best to suppress excited-state effects on the isovector electromagnetic form factors of the nucleon in order to better understand their impact on a possible resolution of the puzzle.

The $eN$ scattering cross section is usually parameterized in terms of the Sachs form factors $G_E$, $G_M$, whereas the matrix element of the vector current between nucleon states has a natural decomposition in terms of the Dirac and Pauli form factors
$F_1$ and $F_2$,
\begin{equation}
\langle N(p',s')|V_\mu|N(p,s)\rangle = 
\overline{u}(p',s')\left[\gamma_\mu F_1 + i\frac{\sigma_{\mu\nu}q_\nu}{2m_N}F_2\right]u(p,s)\,.
\end{equation}
The relationship between the two sets of form factors is given by
\begin{equation}
G_E(Q^2)=F_1(Q^2)-\frac{Q^2}{4m_N^2}F_2(Q^2),~~~G_M(Q^2)=F_1(Q^2)+F_2(Q^2).
\end{equation}

In order to measure the nucleon form factors on the lattice, we compute the two- and three-point functions of nucleon interpolating operators $N^\alpha$
\begin{equation}
C_2(\vec{q},t_s) = \sum_{\vec{x}} \rme^{i\vec{q}\cdot\vec{x}} \Gamma_{\alpha\beta} \langle\overline{N}^\beta(t_s,\vec{x})N^\alpha(0)\rangle,~~~~~~~C_{3,\calO}(\vec{q},t,t_s)=\sum_{\vec{x},\vec{y}} \rme^{i\vec{q}\cdot\vec{y}}\Gamma_{\alpha\beta}\langle\overline{N}^\beta(t_s,\vec{x})\calO(t,\vec{y})N^\alpha(0)\rangle,
\end{equation}
where $\Gamma=\frac{1}{2}(1+\gamma_0)(1+i\gamma_5\gamma_3)$ is a polarization matrix, from whose ratios
\begin{equation}
R_{V_\mu}(\vec{q},t,t_s) = \frac{C_{3,V_\mu}(\vec{q},t,t_s)}{C_2(\vec{0},t_s)} \sqrt{\frac{C_2(\vec{q},t_s-t)C_2(\vec{0},t)C_2(\vec{0},t_s)}{C_2(\vec{0},t_s-t)C_2(\vec{q},t)C_2(\vec{q},t_s)}}
\end{equation}
we can extract the Sachs form factors as
\cite{Jager:2013kha}
\begin{align}
{\rm Re}\left[R_{V_0}(\vec{q},t,t_s)\right] &= \sqrt{\frac{m_N+E_{\vec{q}}}{2E_{\vec{q}}}} G^{\rm eff}_E(Q^2,t,t_s)\\
{\rm Re}\left[R_{V_i}(\vec{q},t,t_s)\right]_{i=1,2} &= \epsilon_{ij}q_j\frac{1}{\sqrt{2E_{\vec{q}}(m_N+E_{\vec{q}})}} G^{\rm eff}_M(Q^2,t,t_s).
\end{align}
Similar relations hold for the axial and induced pseudoscalar form factors
\cite{Capitani:2012gj,LAT14:Parry}.

\section{Excited-State Contributions}

\noindent For $Q^2\not=0$, the excited states contributing to the incoming and outgoing nucleon channels are different in our kinematics (where the final nucleon is at rest), and ignoring finite-size effects we can parameterize the leading excited-state contaminations by
\begin{equation}\label{eq:excstateform}
G^{\rm eff}_X(Q^2,t,t_s) = G_X(Q^2) + c_{X,1}(Q^2)\rme^{-\Delta t} + c_{X,2}(Q^2)\rme^{-\Delta' (t_s-t)} + \ldots
\end{equation}
Since our aim is the extraction of the ground-state form factor $G_X(Q^2)$, we need to find means to identify and remove, or else to suppress in some other way, the leading excited-state contributions. For this end we have considered three different methods
\cite{Capitani:2012ef, Capitani:2012uca}:
\begin{figure}
\begin{center}
\includegraphics[width=0.5\textwidth,keepaspectratio=]{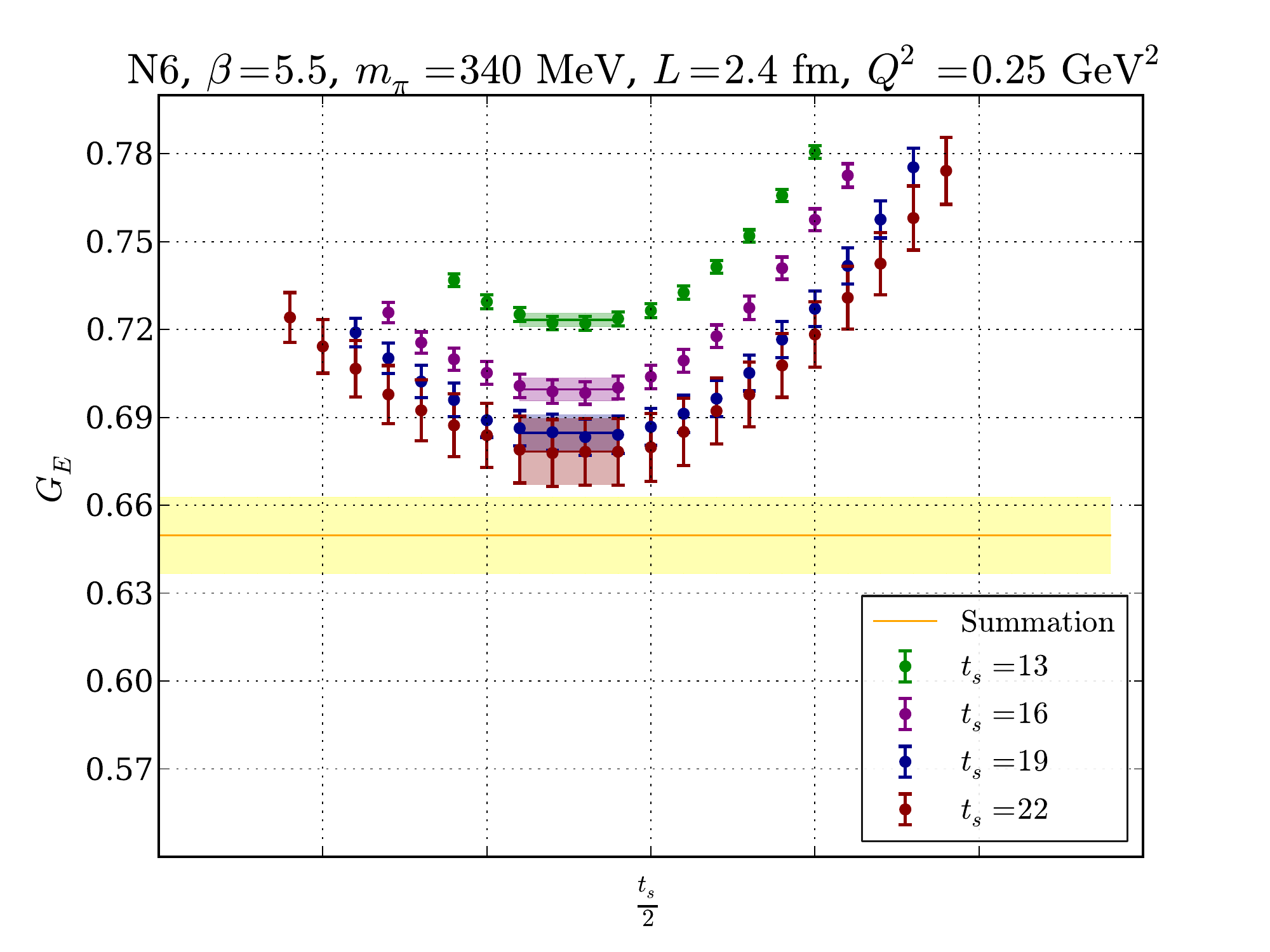}%
\includegraphics[width=0.5\textwidth,keepaspectratio=]{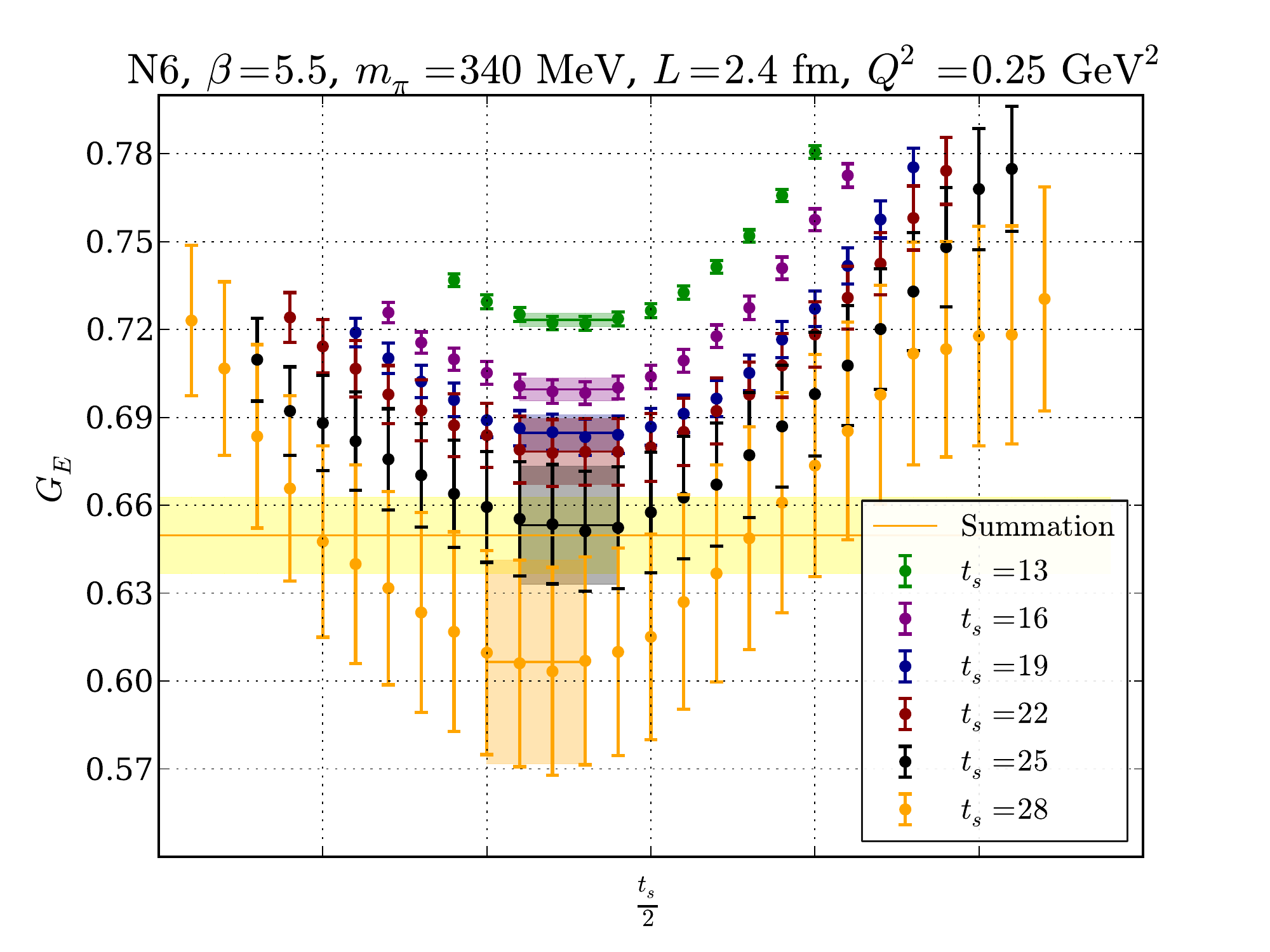}
\end{center}
\caption{Results for $G_E$ at a number of different source-sink separations. Left panel: $t_s\le 1.1$~fm, right panel: $t_s\le 1.4$~fm. Shown are data points and plateaux (as shaded bands); the result from the sumamtion method is shown as the yellow band.}
\label{fig:plateauxN6}
\end{figure}
\begin{itemize}

\item The well-known {\em plateau method} is based on the assumption that excited-state contributions in (\ref{eq:excstateform}) are sufficiently suppressed, and proceeds by looking for a region in $t$ where the signal for $G^{\rm eff}_X(Q^2,t,t_s)$ for given $Q^2$ and $t_s$ has reached a plateau, which is then fitted to a constant identified as $G_X(Q^2)$. While well-tried and simple, this method is beset with a number of problems, the most prominent of which is that in order for the plateau to be a good estimate for $G_X(Q^2)$, the excited-state contributions from both sides need to have decayed by at least one or two e-foldings; in order to achieve this, large values for $t_s$ are required. Unfortunately, the signal-to-noise ratio becomes rather poor as $t_s$ is increased, rendering this difficult to achieve with present statistics. As we observe a systematic trend in the plateau values as a function of $t_s$ even for relatively large distances $t_s\sim 1.4$~fm (cf.~fig.~\ref{fig:plateauxN6}), the practicability of this method is seriously in doubt.

\item The by-now also well-established {\em summation method}
\cite{Maiani:1987by,Capitani:2012gj}
proceeds by forming summed ratios
\[
S_X(Q^2,t_s) = \sum_{t=0}^{t_s} G^{\rm eff}_X(Q^2,t,t_s) \rightarrow c + t_s\left\{G_X(Q^2) + \calO\left(\rme^{-\Delta t_s}\right)\right\}
\]
and fitting their $t_s$-dependence to a straight line, from the slope of which the desired ground-state form factor $G_X(Q^2)$ can be extracted. The great advantage of the summation method is that it leads to a parametrically reduced excited-state contamination, since any $t$-dependence has been eliminated by summing over $t$, and $\Delta t_s\gg \Delta t$ by construction. The summation method's major disadvantage is a significant increase in statistical errors when compared to the plateau method. One could therefore argue that the reduction in (hard to control) systematic error is paid for by a corresponding increase in (expensive to reduce) statistical error.

\item To have a further handle on the precise behaviour of the excited-state contributions, we have also considered {\em two-state fits}
\cite{Capitani:2012ef,Green:2011fg},
i.e. we have explicitly fit $G^{\rm eff}_X(Q^2,t,t_s)$ to leading excited-state contributions of the form (\ref{eq:excstateform}) with $\Delta=m_\pi$, $\Delta'=2m_\pi$, both
as a function of $t$, $t_s-t$ at each $t_s$ separately, and
as a function of $t_s$, $t$ at all $t_s$ simultaneously.
The advantage of these fits is that they are able to fully remove the leading excited-state contamination contained in the signal and project onto the ground state without any need to further increase the range of $t$ and $t_s$ under consideration. Their disadvantage lies in that they are somewhat dependent on our assumptions about the excited states (in particular the energy gap), making it somewhat difficult to assess the overall trustworthiness of the fitted results. In light of this latter disadvantage, we do not currently expect to quote final results from the two-state fits, but only to use them as a control to verify that excited-state contaminations do not seriously affect our final answers from the other methods.

\end{itemize}

\begin{figure}
\begin{center}
\includegraphics[width=0.5\textwidth,keepaspectratio=]{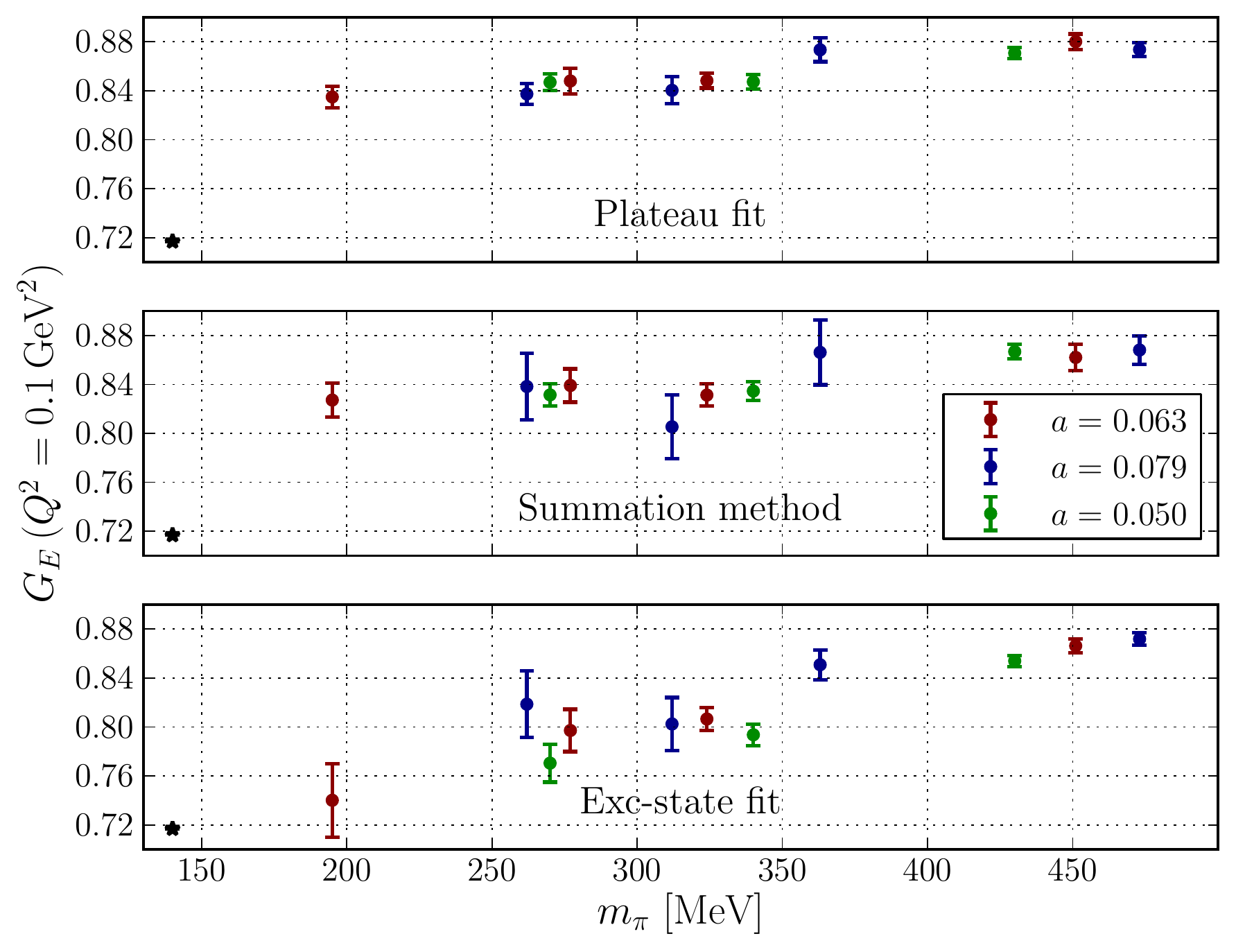}
\end{center}
\caption{Comparison of the chiral behaviour of (from top to bottom) the plateau method, summation method and simultaneous two-state fits. For details cf.~text.}
\label{fig:cmpmethods}
\end{figure}

In fig.~\ref{fig:cmpmethods}, we compare the pion-mass dependence of the value of the electric Sachs form factor evaluated at $Q^2=0.1$~GeV$^2$, $G_E(0.1~\textrm{GeV}^2)$, as determined using the different methods. For the plateau method, there appears to be no appreciable pion-mass dependence, and the chiral extrapolation does not approach the experimental value as the pion mass is reduced. The summation method shows a slight improvement in so far as a chiral trend is visible, but an extrapolation would still miss a experimental value by a significant margin (although the discrepancy is less significant than for the plateau method, a large portion of this is due to the larger errors associated with the summation method). By contrast, the simultaneous two-state fits yield results that show a clear chiral trend and point towards the experimental value in the physical limit.

\begin{figure}
\begin{center}
{\includegraphics[width=0.5\textwidth,keepaspectratio=]{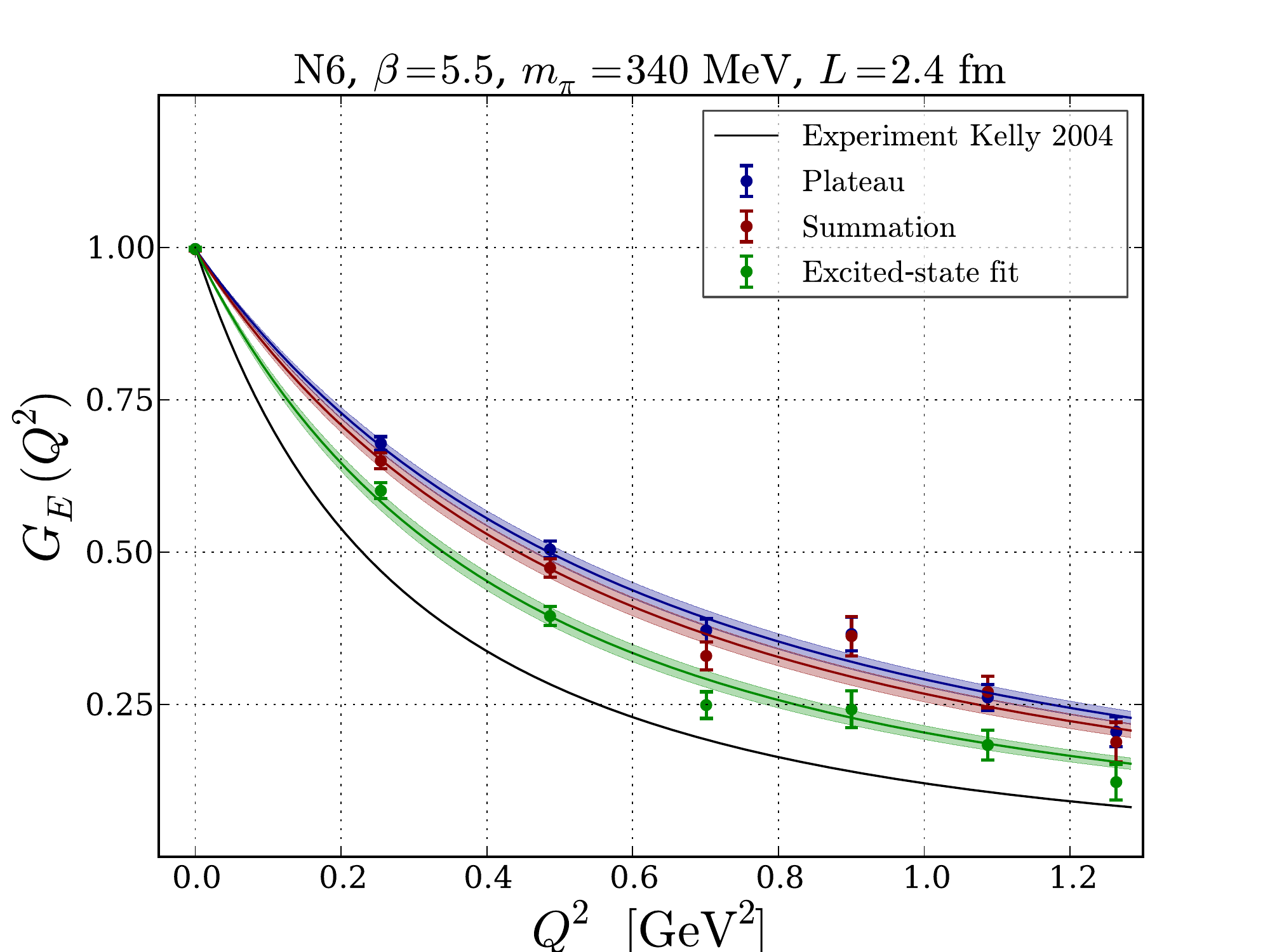}}{\relax}%
{\includegraphics[width=0.5\textwidth,keepaspectratio=]{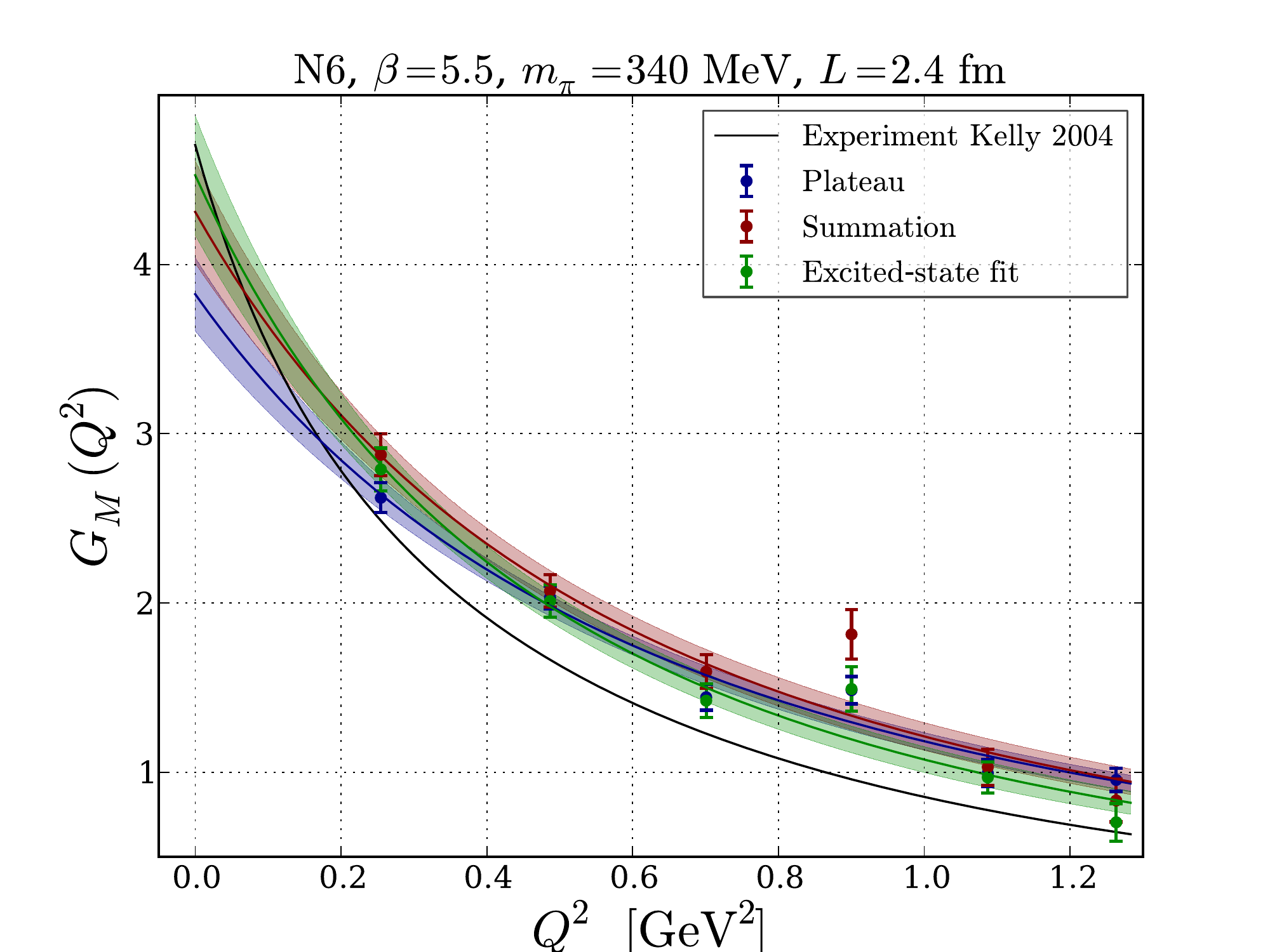}}{\relax}%
\end{center}
\caption{The Sachs form factors $G_E$ (left) and $G_M$ (right) as a function
of $Q^2$ on one of our ensembles.
The dipole fits are shown as shaded bands; also show for comparison as a black line is the Kelly parameterization of the experimental results
\cite{Kelly:2004hm}.
}
\label{fig:ffplots}
\end{figure}

\noindent By parameterizing each Sachs form factor as a dipole
\begin{equation}
G_X(Q^2) = G_X(0)\left(1+\frac{Q^2}{M_X^2}\right)^{-2}
\end{equation}
where $G_E(0)=1$, $G_M(0)=\mu$, we can extract the charge radii and anomalous magnetic moment $\kappa$ via
\begin{equation}
\frac{1}{M_E^2} = \frac{r_E^2}{12} = \frac{r_1^2}{12}+\frac{\kappa}{8m_N^2} ~~~~~~~~~~
\frac{1}{M_M^2} = \frac{r_M^2}{12} = \frac{r_1^2+\kappa r_2^2}{12(1+\kappa)}
\end{equation}
where $\kappa=\mu-1$.

\begin{figure}
\begin{center}\vskip-5ex
\includegraphics[height=0.5\textheight,keepaspectratio=]{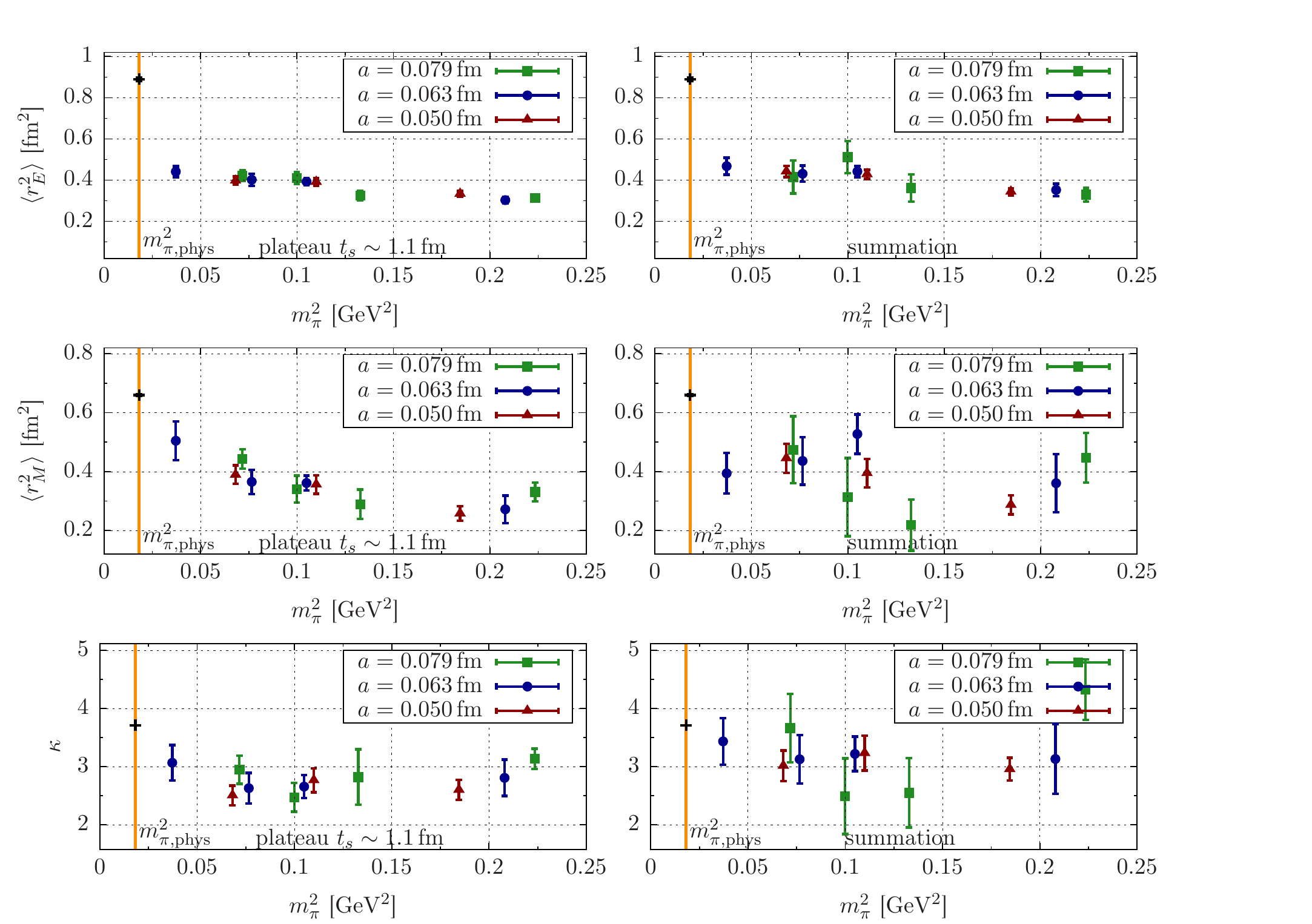}
\end{center}
\caption{The charge radii and anomalous magnetic moment of the nucleon as a function of the pion mass. Left: plateau method ($t_s\sim 1.1$~fm), right: summation method ($t_s\le 1.1$~fm). Also shown for comparison are the experimental values and the physical pion mass.}
\label{fig:radcomp}
\end{figure}

Fig.~\ref{fig:ffplots} shows the data for $G_E$ and $G_M$ together with the dipole fits. The pion-mass dependence of the results obtained for the radii and the magnetic moment from the plateau and summation methods, respectively, is shown in fig.~\ref{fig:radcomp}.

\section{A Possible Scenario}

\noindent In order to explain why we observe such a different chiral behaviour for the plateau and summation methods on the one hand, and the two-state fits on the other hand, it helps to consider a scenario in which we assume the two-state fits to be a true description of the data. Under this assumption, we can reconstruct the summed ratios that we would obtain at various value of $t_s$, and compare the slope obtained from fitting a straight line to the reconstructed values with the ``true'' value of the form factor.

\begin{figure}
\begin{center}
\includegraphics[width=0.5\textwidth,keepaspectratio=]{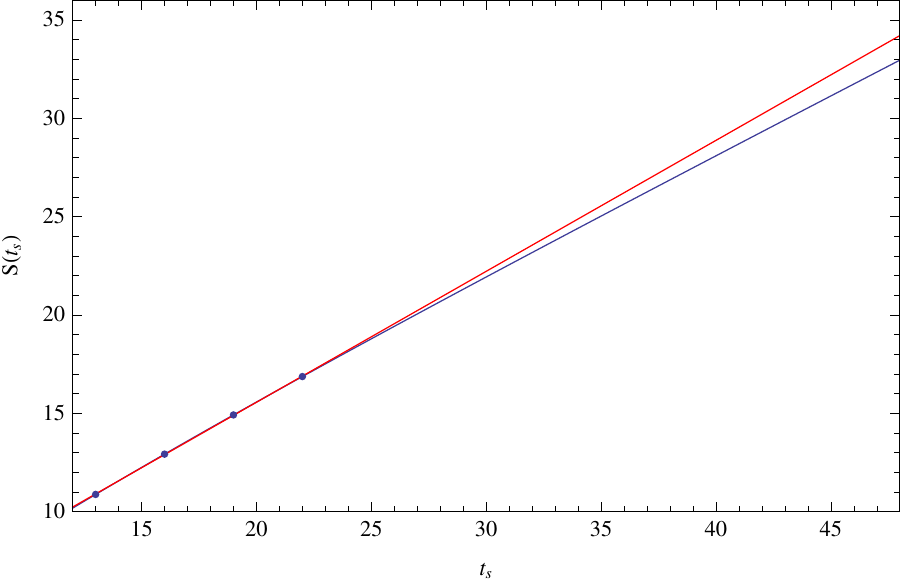}~\includegraphics[width=0.5\textwidth,keepaspectratio=]{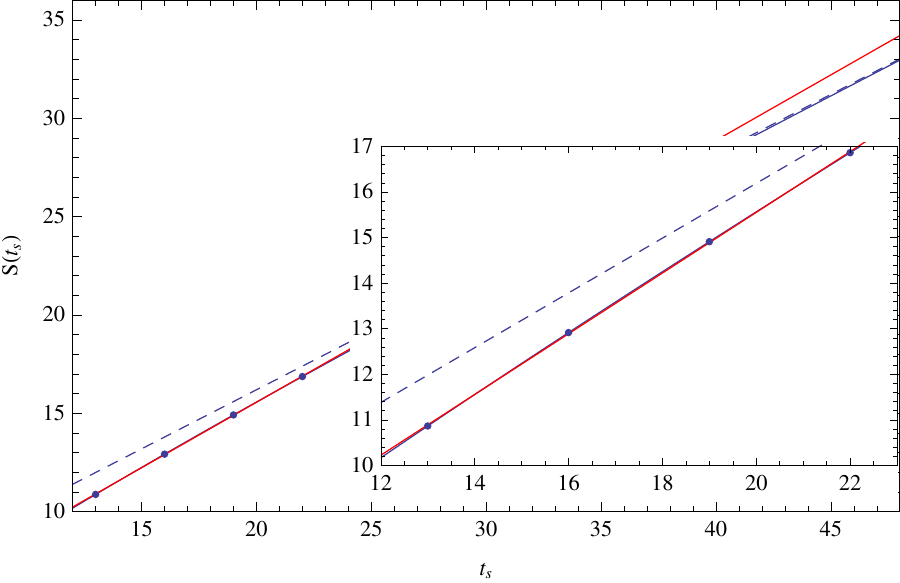}
\end{center}
\caption{A comparison between the linear fit (red line) to summed ratios reconstructed from two-state fits (blue points and line) and the correct asymptotic behaviour (dashed blue line) inferred from the two-state fits. All numbers are given in lattice units ($a=0.050$~fm). For details, cf.~the text.}
\label{fig:scenario}
\end{figure}

An example of such a reconstruction on one of our ensembles is shown in fig.~\ref{fig:scenario}. In the left panel, we show the reconstructed ratios for the values of $t_s$ considered here as blue dots, with the straight-line fit to those reconstructed points shown in red. Also shown as a blue line is the functional form of the summed ratios as a function of $t_s$ over a much larger range of source-sink separations. It can be seen that the slope of the red line does not agree well with that of the blue curve as $t_s$ grows. In the right panel, we have added the asymptotic linear behaviour of the summed ratio (the dashed blue line). As can be seen from the inset, while the straight-line fit to the reconstructed ratios is indistinguishable from the true functional form at low values of $t_s$, it is quite different from the correct asymptotic value.

This behaviour is to be contrasted with the one observed for $g_A$
\cite{Capitani:2012gj},
where the summation method has been found to be very effective in extracting the asymptotic behaviour. A crucial difference between the two cases is that for the form factors at $Q^2\not=0$, the gap appearing in the leading exponential corrections to the summed ratio is only $m_\pi$, rather than $2m_\pi$ as for the $Q^2=0$ case. This makes the approach to the asymptotic linear behaviour much slower, and (at least at first sight) seems to require twice as large a source-sink sparation $t_s$ in order to obtain the same reduction in systematic error.

\section{Conclusions}

\noindent A systematic trend in the value of $G_E$ obtained from the plateau method persists to $t_s\sim 1.4$~fm, and even with the summation method, $G_E$ systematically comes out too high; considering only the largest values of $t_s$ in the summation method brings the results closer to experiment, albeit at the expense of very large statistical errors.

The impact of excited states on the results of the plateau, and to a lesser extent also the summation method, appears to increase even further as the chiral limit is approached.

A possible reason for these observations is indicated by two-state fits: Given the appearance of the small gap $m_\pi$ for $Q^2\not=0$, the approach to the plateau is very slow, and the summed ratios still receive sizeable corrections to their asymptotic linear behaviour.\\

{\noindent {\bf Acknowledgments:}
We are grateful to our colleagues within the Coordinated Lattice Simulations
(CLS) initiative for sharing ensembles.
This work was granted access to the HPC resources of the Gauss Center for
Supercomputing at Forschungzentrum Jülich, Germany, made available within
the Distributed European Computing Initiative by the PRACE-2IP, receiving
funding from the European Community’s Seventh Framework Programme
(FP7/2007-2013) under grant agreement RI-283493 (project PRA039).
We are grateful for computer time allocated to project HMZ21 on the
JUQUEEN BG/Q computer at NIC, Jülich.
This work was supported by the DFG via SFB~1044
and grant HA~4470/3-1. We thank Dalibor Djukanovic and Christian Seiwerth
for technical support.
}

\end{document}